# Establishing a Multi-Thesauri-Scenario based on SKOS and Cross-Concordances

Philipp Mayr
GESIS – Leibniz Institute for the Social Sciences, Germany
philipp.mayr@gesis.org

Benjamin Zapilko
GESIS – Leibniz Institute for the Social Sciences, Germany
benjamin.zapilko@gesis.org

York Sure
GESIS – Leibniz Institute for the Social Sciences, Germany
york.sure@gesis.org

**Keywords:** SKOS; thesauri; cross-concordances; interoperability; Linked Open Data

## Introduction

With the standardization of SKOS[1] (Simple Knowledge Organization System) in August 2009 a data model has been offered to publish controlled vocabularies and taxonomies on the web in a technical and semantically interoperable way. The heterogeneous environment of various vocabularies worldwide can be technically harmonized prospectively and especially the content of traditional databases can be made accessible and connectable for applications of the Semantic Web, i.e. as Linked Open Data[2]. Vocabularies in SKOS format and respectively crosswalks between them can play a relevant role in this context, because they can serve as a bridging hub for the inter-linking of different published and indexed data sets.

This case study proposes a scenario with three topic-related thesauri, which have been connected with bilateral cross-concordances as part of a major terminology mapping initiative in the project KoMoHe (Mayr & Petras, 2008). The thesauri have already been or will be converted to SKOS and in order to not omit the relevant crosswalks, the mapping properties of SKOS will be used for modeling them adequately. The participating thesauri in this approach are: (i) TheSoz (Thesaurus for the Social Sciences, GESIS) which has been converted to SKOS in a first experimental version (Zapilko & Sure, 2009) in 2009, the current version uses SKOS-XL for representing preferred and non-preferred terms and defines additional extensions which are oriented on the introduced SKOS extensions of the EUROVOC thesaurus (Smedt, 2009) to model more complex relations between terms, i.e. "use combination" relations, (ii) STW (Standard-Thesaurus for Economics, ZBW) which has also been published in SKOS format (Neubert, 2009) and (iii) IBLK-Thesaurus (SWP).

Currently, the conversion of vocabularies to SKOS is an active research area, but there are still unsolved and relevant issues which could not be treated satisfyingly yet. Our approach focuses on the application of existing crosswalks to the SKOS mapping properties and the establishment of a linked data application based on those connected thesauri.

## Modeling Cross-Concordances in SKOS

The SKOS mapping properties provide standardized relations in order to link SKOS concepts of different concept schemes, which are represented in this scenario by three participating thesauri. When modeling cross-concordances in SKOS format inconsistencies and problems can occur which are caused by idiosyncrasies in thesauri. Although SKOS provides a standard model for representing vocabularies, transformed or converted thesauri can differ a lot due to various complexity and heterogeneous structure. Modeling mostly term-based thesauri in a concept-based way can be realized differently. A reason for inconsistencies is that the given cross-concordances where defined on term-based thesauri, but the SKOS versions of those thesauri are concept-based.

---

[1] http://www.w3.org/2004/02/skos/
[2] http://linkeddata.org/

Therefore crosswalks between traditional thesauri cannot simply be adapted to the SKOS mapping properties under certain conditions. It has to be proven if the two terms of a given crosswalk represent adequate concepts in the corresponding SKOS versions by i.e. being used as skos:prefLabel in a concept. In case of the cross-concordances defined in the KoMoHe project they were only defined between preferred terms which means that a conversion to SKOS should be feasible without further complications. In general, if the above described requirements are met, existing cross-concordances can be relatively easy be transformed to SKOS (see listing 1). Depending on where the SKOS cross-concordances are physically stored the full URIs of the references concepts have to be addressed.

```
<rdf:Description rdf:about="http://lod.gesis.org/thesoz/concept/10039068">
      <skos:exactMatch rdf:rescource="http://zbw.eu/stw/descriptor/11971-0">
</rdf:Description>
```

Listing 1: Example for a simple cross-concordance in SKOS between TheSoz and STW

For the case that there are crosswalks between non-preferred terms, each participating SKOS vocabulary has to be checked on how non-preferred terms are modeled, because the mapping properties of SKOS can only be used between concepts. At current state those crosswalks could not directly be modeled in SKOS, additional extensions would have been to define in order to preserve the relevant information.

Domain-specific differences in thesauri can cause conversion problems either. For example, a concept in one thesaurus can correspond to a combination of two concepts in another thesaurus. Cross-concordances can be in such a complex manner like associate relations between terms of one vocabulary. But the mapping properties of SKOS are too restrictive in their current definition that alternative possibilities, i.e. defining own extensions, have to be defined on how to deal with these special use cases.

## Establishing a Multi-Thesauri-Scenario as a Linked Data Application

In order to provide interoperability between the participating thesauri and the external data sets, the thesauri and the cross-concordances between them have to be made accessible on the web in an integrated way. The most common solution is to publish them via multiple SPARQL[3] endpoints according to assumed physically different storage locations (thesauri of different organizations). A linked data interface, i.e. the Pubby[4] linked data frontend, is set upon these endpoints to generate a combined html representation of the different thesauri via dereferencing the URIs of the participating thesauri and their cross-concordances. With the inclusion of crosswalks a stronger interlinking between the thesauri is represented which is not only based on a term or lexical level, i.e. established via owl:sameAs, but also on precise mappings between the concepts.

## Conclusion and Further Research

The proposed case study outlines the potential of thesauri, which are bi-directional connected via cross-concordances, to serve as a bridge for a stronger interlinking of data sets on the web, i.e. in the manner of the idea of Linked Open Data. Transforming existing vocabularies and thesauri to SKOS remains a complex issue according to the heterogeneous structure of the involved vocabulary. Especially the SKOS conversion of given crosswalks which have a term-based origin can bear major problems when the participated terms are not preferred terms which would usually be represented as concepts in SKOS. In that case and in case of the requirement of semantically more complex relations, i.e. "use combination" relations between terms of different vocabularies, extensions have to be defined if the relevant information of the crosswalks should be preserved.

---

[3] http://www.w3.org/TR/rdf-sparql-query/
[4] http://www4.wiwiss.fu-berlin.de/pubby/

One of the next steps in establishing such a proposed multi-thesauri-scenario is the technical implementation as a test application. Furthermore a generic method has to be developed on how to convert existing cross-concordances to SKOS with a minimal effort in intellectual post-processing.